\documentclass[journal]{IEEEtran}
\usepackage{graphicx}
\usepackage{amsmath}
\usepackage{amssymb}

\begin{document}

\title{GePEToS : A Geant4 Monte Carlo simulation package for Positron Emission Tomography}

\author{S\'ebastien~Jan,
        Johann~Collot,
	Marie-Laure~Gallin-Martel,
	Philippe~Martin,
	Fr\'ed\'eric~Mayet,
	Edwige~Tournefier
\thanks{S. Jan, J. Collot, M.L. Gallin-Martel, Ph. Martin, F. Mayet, E. Tournefier are with the Laboratoire de Physique Subatomique et de Cosmologie, 53 avenue des
Martyrs, 38026 Grenoble cedex France.}
\thanks{S. Jan, present adress : CEA, Service Hospitalier Fr\'ed\'eric Joliot, 4 place G\'en\'eral Leclerc 91406 Orsay France.}
\thanks{E. Tournefier, present adress : Laboratoire d'Annecy-le-vieux de Physique des Particules, 9 Chemin de Bellevue, BP 110 74941 Annecy-le-Vieux cedex
France.}}

\markboth{IEEE TRANSACTIONS ON NUCLEAR SCIENCE}{}

\maketitle

\begin{abstract}GePEToS is a simulation framework developed
 over the last few years for assessing the 
 instrumental performance of future PET scanners.
 It is based on Geant4, written in Object-Oriented C++ and runs on Linux platforms.
 The validity of GePEToS has been tested on the well-known Siemens ECAT EXACT HR+ camera.
 The results of two application examples are presented : the 
 design optimization of a liquid Xe $\mu$PET camera dedicated
 to small animal imaging as well as the
 evaluation of the effect of a strong axial magnetic field on the image 
 resolution of a Concorde P4 $\mu$PET camera. 
\end{abstract}


\section{Introduction}

Over the last decade, the performance of Positron Emission Tomography (PET) scanners
 have considerably improved. 
For instance, commercial cameras dedicated to small animals now feature a space 
resolution below 2~mm along with a sensitivity better than 1\%~\cite{bib:P4}. 
No matter how beneficial this performance gain has been to users, for instrument designers,  
it has somehow hardened the challenge of finding new solutions 
which would go beyond the present instrumental limits at an affordable cost. 
This is why the complete simulation of new TEP configurations 
under study has now become even more 
justified than in the past, and calls for the development of a dedicated simulation 
framework, sufficiently versatile to allow fast and very detailed approaches with
the best-existing emulation of all the underlying physical processes.
Since its first public release in 1998, the stability, the validity and hence the popularity 
of Geant4~\cite{bib:G4}- the Object-Oriented particle tracking and transport simulation framework developed
by the High Energy Physics community - have noticeably progressed. 
In our opinion, it has  
become the best toolkit from which any common and public TEP simulation framework 
should be developed.
GePEToS : {\bf Ge}ant4 {\bf P}ositron {\bf E}mission {\bf To}mography {\bf S}imulation, 
is a first attempt that we have made over the last few years to reach this goal .

\section{Adequacy for TEP scanners of Geant4-simulated physical processes}
\begin{figure}
\centering
\includegraphics[width=3.5in]{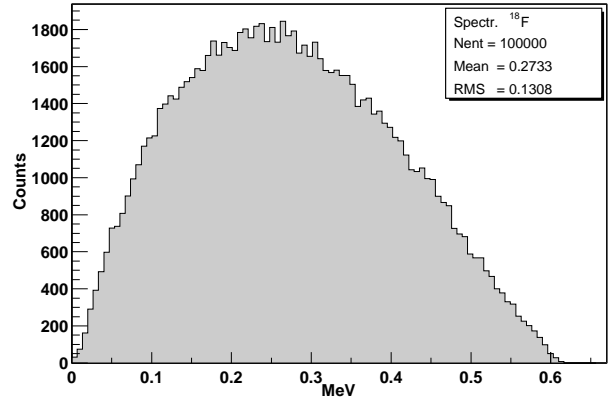}
\caption{$\beta^{+}$ spectrum for $^{18}$F as produced by GePEToS.}
\label{enerf18}
\end{figure}

As we aimed at providing the possibility to fully simulate all the processes 
taking place during the short life of a positron and then in the 
transport and interaction of its two 511~keV annihilation photons, efforts were made to check the validity of Geant4 and if needed corrections or modifications were done.
Hence, we have added as part of GePEToS the possibility to generate positrons
according to their respective $\beta^{+}$ spectra for $^{18}$F, $^{15}$O, $^{11}$C (figure~\ref{enerf18} for $^{18}$F).
Table \ref{tab:ener} shows the good agreement between our simulation and experimental data for the maximum and the most probable energies for each spectrum.
After this step, the positrons are fully tracked down until they annihilate. 

\begin{table}[htbp]
\begin{center}
\begin{tabular}{|c|c|c|c|c|} \hline
&\multicolumn{2}{|c|}{$E_{max}$} & \multicolumn{2}{|c|}{$<E>$ : Most Probable Energy}  \\
 \cline{2-5}
         & \# Simulation & \# Data~\cite{table} & \# Simulation  & \# Data~\cite{table}    \\\hline 
$^{18}F$ & 620 keV & 633 keV  &  250 keV  & 242 keV \\\hline 
$^{11}C$ & 950 keV & 959 keV  & 375 keV  & 385 keV \\\hline 
$^{15}O$ & 1750 keV & 1738 keV &  725 keV  & 735 keV \\ \hline\hline 
\end{tabular}
\end{center} 
\caption{\it $\beta^{+}$ spectra comparisons between simulation and experimental data~\cite{table}.}
\label{tab:ener}
\end{table}

\begin{figure}
\centering
\includegraphics[width=3.5in]{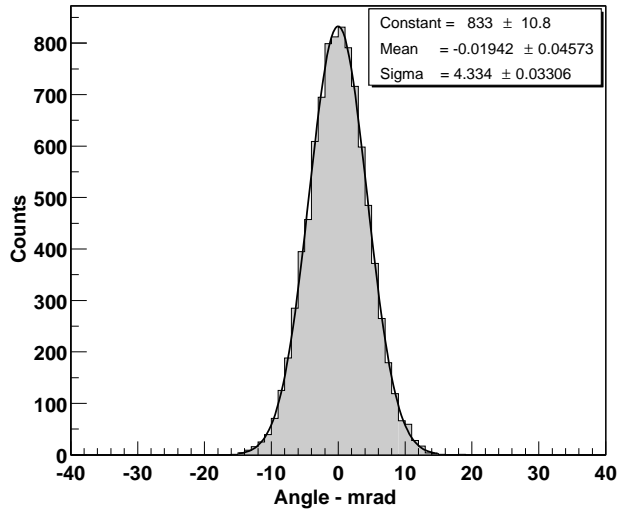}
\caption{Acolinearity angle distribution in water between the two annihilation $\gamma$ after correction in Geant4}
\label{aco}
\end{figure}

As Geant4 did not correctly reproduce the acolinearity of the two annihilation photons 
(($\theta_{\gamma\gamma} - \pi$ rad.) $\simeq$ 10 mrad FWHM) which affects the image 
resolution, the native Geant4 algorithm has been modified to obtain a correct annihilation behavior in water where this phenomenon takes place in PET (figure ~\ref{aco}). 
In water, the experimental measurement shows an energy shift between the two annihilation photons due to the orbital motion of the atomic electrons. This energy distribution is
gaussian centered on zero with $\Delta E = 2.59$ keV (FWHM)
~\cite{posit}. The relation between the energy shift and the acolinearity angle distribution is given by the formula :
$$
\theta_{\gamma\gamma} = \frac{2.\Delta E}{m_{o}.c^{2}}
$$

\begin{figure}
\centering
\includegraphics[width=3.5in]{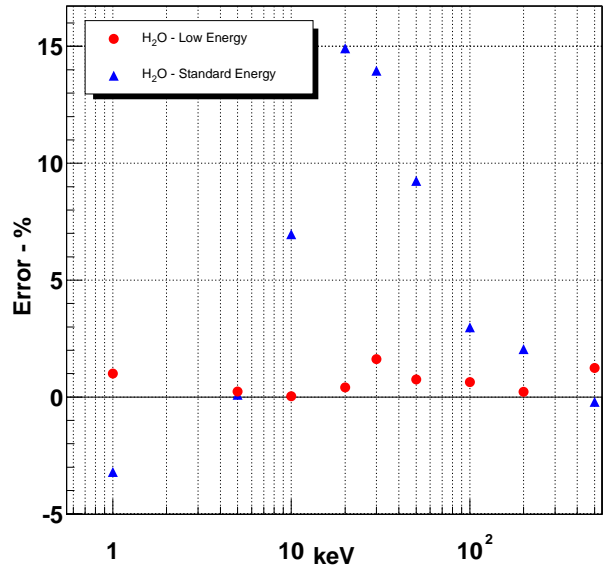}
\caption{Deviation of the photon total attenuation coefficients as computed by Geant4 
from NIST the data for two Geant4 EM process packages.}
\label{LEdata}
\end{figure}

Finally, two Geant4 electromagnetic (EM) process packages have been tested : the Standard one and the LowEnergy one, dedicated to low EM physical processes. Comparisons 
between these two packages and the NIST experimental data ~\cite{bib:NIST} have been achieved. As an example, results on the total attenuation coefficient are presented
hereafter on figure~\ref{LEdata} : the low energy process package provides results in much better agreement with the experimental data.
The difference is explained by the absence of Rayleigh scattering in the standard EM
package. Indeed, the Low Energy package is now used in GePEToS even though it slightly increases the code CPU consumption.    

\section{Framework description}

GePEToS as Geant4, is fully written in Object-Oriented C++. It runs on 
Linux platforms (tested on RedHat 6.2). 
It uses a simple mechanism to define the geometry and the material composition 
of most of the PET cameras currently commercialized or under development
(multi-ring and multi-crystal block devices). This is achieved
by setting an ASCII configuration file in which users can select the desired 
$\beta^{+}$ isotope, the number of active rings, the dimensions and the 
segmentation of the crystal blocks, the nature of the crystal (NaI, LSO, BGO), the phantom
type and the acquisition mode (2D or 3D). In addition, users have to provide the energy 
resolution measured or estimated at 511 keV which is then normally scaled 
according to the energy deposited in the crystals. For standard configurations, 
neither code modification nor recompilation are necessary.  
For every positron event and each of its two annihilation photons, 
GePEToS computes the deposited energy and an energy-weighted 
position in the transverse plane of the crystals which is then used to 
determine which crystal was hit. The depth of interaction (DOI) in crystals
is also computed and stored if this readout mode is selected by users.
Hit information is written in ROOT~\cite{bib:ROOT} files. 
Sinograms are separately prepared by using a ROOT application and finally
processed by an IDL program~\cite{bib:IDL} to reconstruct the tomographic images. 
More complex geometries, departing from the standard multi-ring crystal 
block model, can be handled with minor modifications of the code and by rebuilding the application.

\section{Validation tests}

\begin{figure}[t]
\centering
\includegraphics[scale=1]{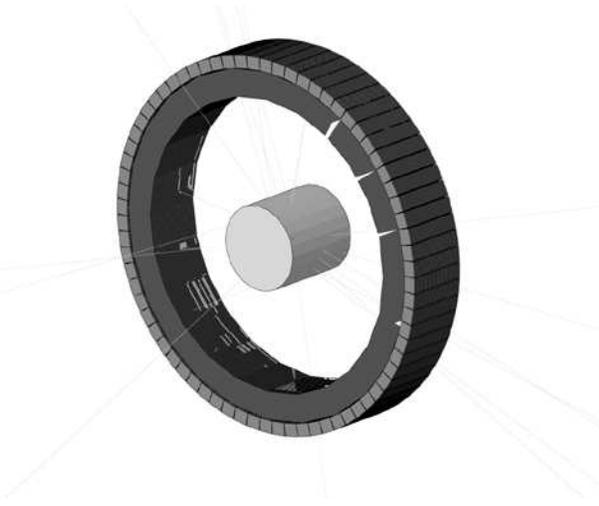}  
\caption{Graphical view of the ECAT EXACT HR+ PET camera as 
described in GePEToS in 2D acquisition mode, with the lead septa 
slid in front of the crystal rings - Rays exiting the 
water phantom represent a few simulated annihilation photons.}
\label{fig:HR}
\end{figure}

\begin{figure}[p]
\centering
\includegraphics[scale=1.4]{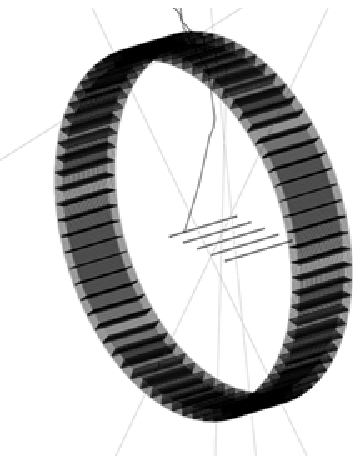}  
\caption{Graphical view of the ECAT EXACT HR+ PET camera as 
described in GePEToS in 3D acquisition mode, without the lead septa - On this example, we used five $^{18}$F-loaded capillaries as phantom to evaluate the image resolution.}
\label{fig:HR3D}
\centering
\includegraphics[width=3.5in]{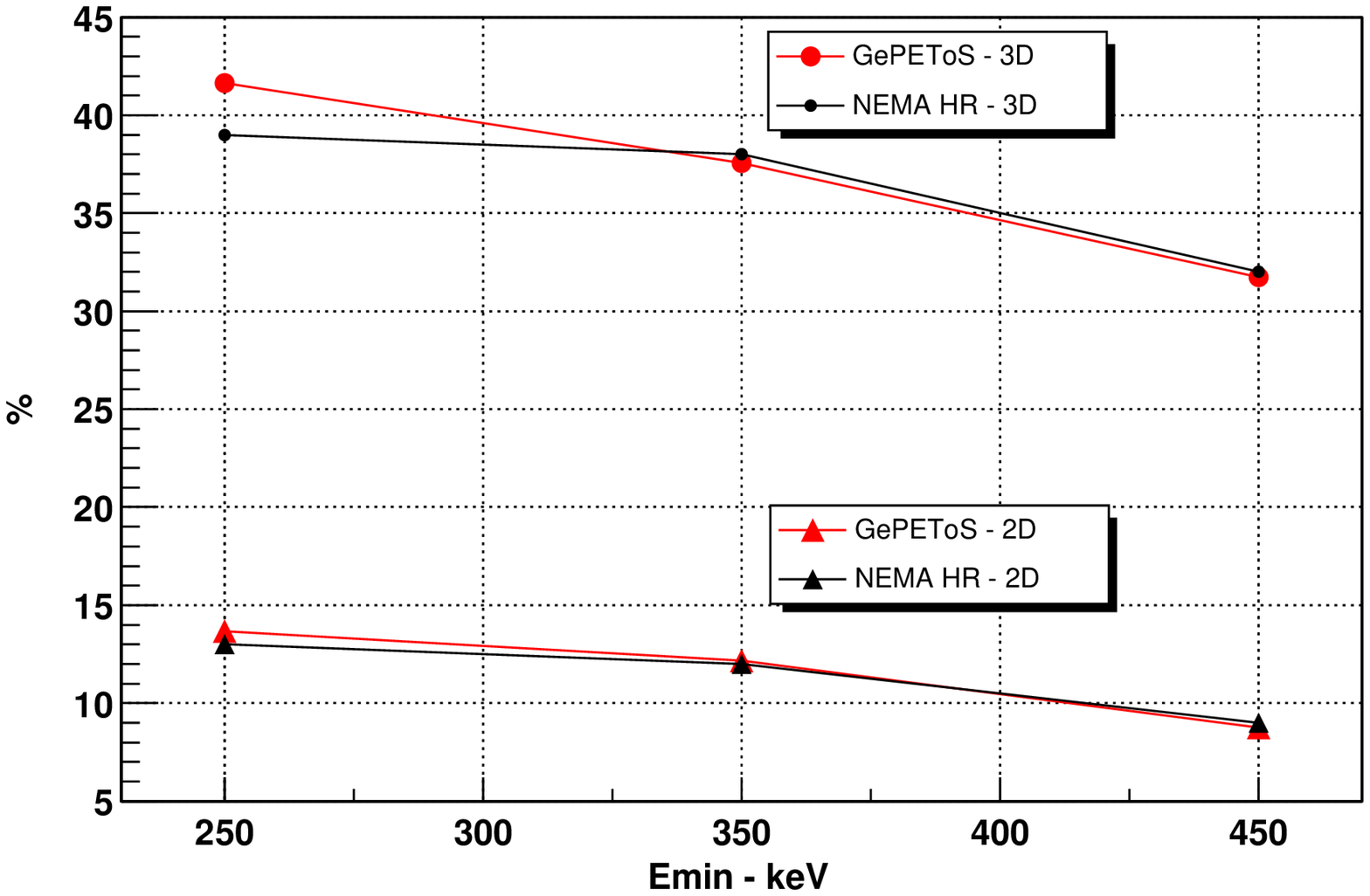}   
\caption{Simulated fraction of scattered coincidences as a function of the energy threshold, 
compared to experimental data obtained following the NEMA 
protocol in 2D and 3D acquisition modes.}
\label{fig:sf_NEMA}
\includegraphics[width=3.5in]{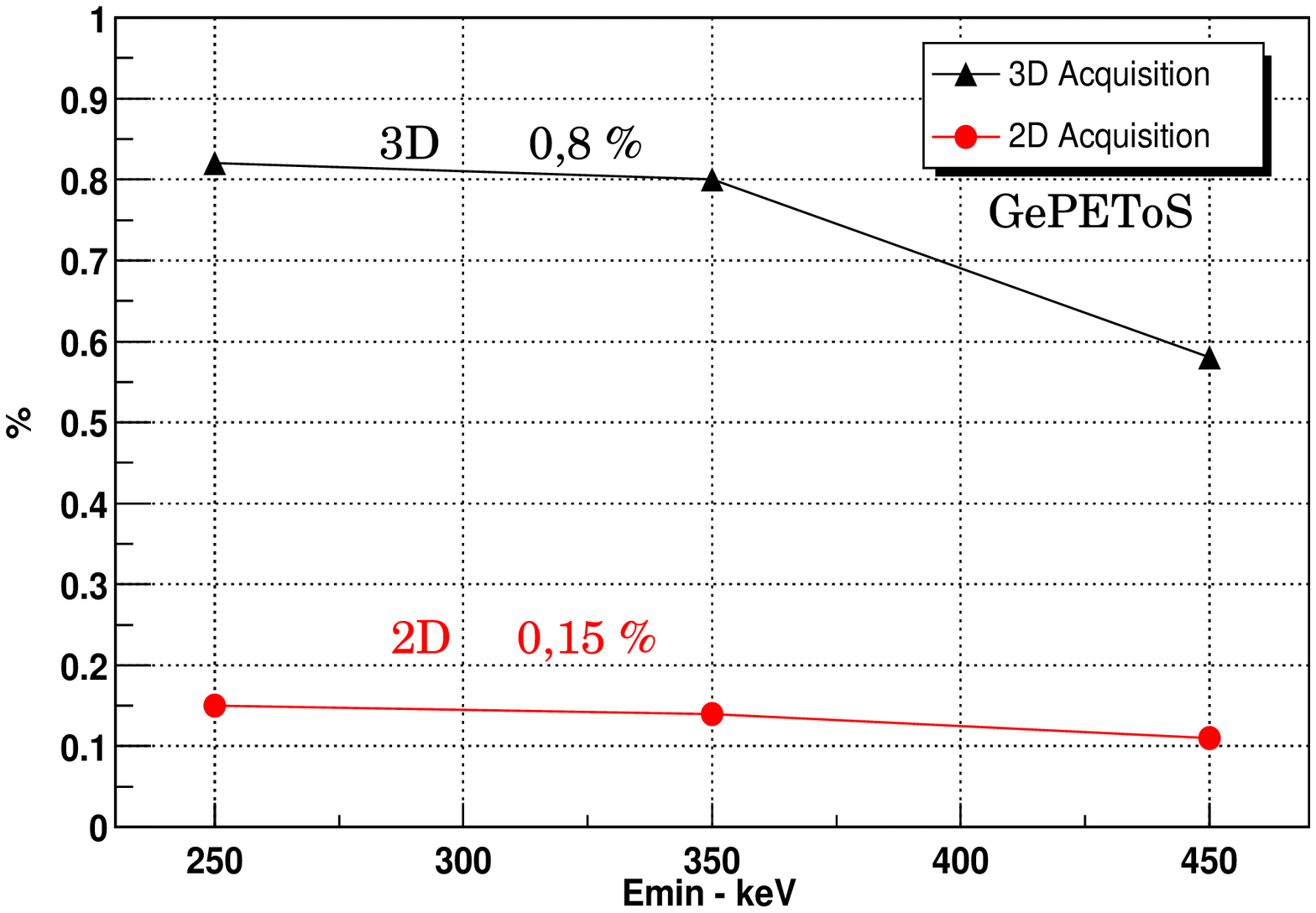}   
\caption{Simulated sensitivity as a function of the energy threshold
obtained following the NEMA 
protocol in 2D and 3D acquisition modes.}
\label{fig:sensi}
\end{figure}

An exhaustive validation test of GePEToS has been achieved on one of the most common
PET cameras used in medical examination centers : the Siemens ECAT EXACT HR+ PET scanner~\cite{bib:ECAT}. 
The ECAT EXACT HR+ PET scanner consists of 32 rings, featuring an internal diameter of 82.7~cm and spanning
15.2~cm in the axial field of view. It is made of blocks of BGO crystals. Each crystal has
a transverse cross-section of 4 x 4.1 mm$^2$ and is 30 mm long. This device, as modeled in 
GePEToS, is presented on figure~\ref{fig:HR}, prepared for a 2D acquisition, for which the lead septa have 
been slid in front of the crystal rings. Also shown on the picture is one of the typical 
water phantoms ($\Phi$=20~cm, L=20~cm) that can be used in GePEToS to assess 
the performance of the cameras. Figure~\ref{fig:HR3D} also shows the HR+ scanner but in configuration of 3D acquisition mode, with the septa retracted.

All comparisons of the simulated performance against the available experimental data~\cite{bib:DATA} do show an excellent agreement. To illustrate 
this statement, figure~\ref{fig:sf_NEMA} shows the results for the fraction of scattered coincidences as measured in the NEMA~\cite{bib:NEMA} experimental protocol in 2D and 3D 
acquisition modes with $^{18}$F. For the scatter fraction evaluation, the phantom which is defined in the
NEMA protocol is a water cylinder ($\Phi$=20~cm, L=20~cm) with three
axial $^{18}$F-loaded capillaries placed at the center, at 4 cm and 8 cm in the transaxial plane of the cylinder. The evaluation of total scatter fraction is given by
this expression :
$$
SF_{total} = \frac{1}{25}.\left[ SF_{r_{0}}+8.SF_{r_{4}}+16.SF_{r_{8}} \right]
$$ 
Here, $SF_{r_{0}}$, $SF_{r_{4}}$ and $SF_{r_{8}}$ are respectively the scatter fraction at the center, 4 cm and 8 cm in the phantom. The value of the transaxial Field
Of View (FOV) is taken as 25 cm. The scatter fraction is defined for coincidences included in the $\left[E_{min} ; 650 keV\right]$ energy window.

Figure \ref{fig:sensi} presents the simulation results for the sensitivity evaluation. The NEMA protocol for this calculation assumes a water cylinder filled with 
$^{18}$F for the phantom. The sensitivity value is determined by this equation :
$$
S_{ensi} = \frac{N_{\mbox{coinc}}}{N_{\beta^{+}}}(1-SF_{total})
$$   
For low energy cut values (250 keV and 350 keV), the sensitivity is 0.8 \% and 0.15 \% for 3D and 2D acquisition mode respectively. These simulation results can be
compared to experimental values~\cite{bib:ECAT} which produce 0.8 \% in 3D mode and 0.15 \% for a 2D acquisition.

\begin{figure}[t]
\centering
\includegraphics[width=3.5in]{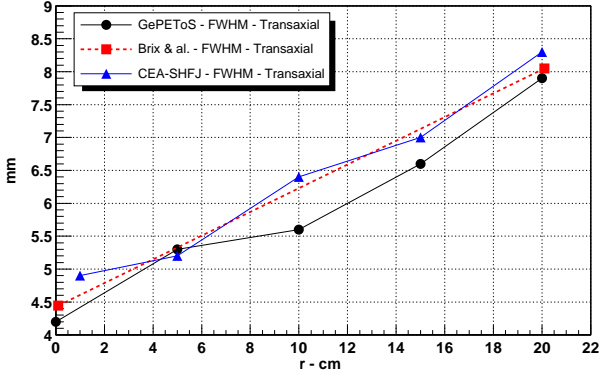}   
\caption{ Radial resolution of the ECAT EXACT HR+ PET camera 
obtained by GePEToS, compared to experimental data.}  
\label{fig:rad_reso}
\end{figure}

Also presented on figure~\ref{fig:rad_reso} is the transaxial image resolution 
determined with five axial $^{18}$F-loaded capillaries (figure~\ref{fig:HR3D}) which again 
shows a good agreement between the simulated and the experimental data. The results are produced by the FWHM gaussian fit on the pixel distribution of the reconstruted image. 
All quantitative results are reported in the table~\ref{tab:resol}.

\section{Application examples}

In this section, we briefly describe two application examples of GePEToS
which show that this simulation framework although in an infant stage,
 can be used to investigate a wide variety of PET problems.

\newpage 
\begin{table}[htbp]
\begin{center}
\begin{tabular}{|c|c|c|} \hline
&\multicolumn{2}{|c|}{Transaxial Image Resolution (FWHM)}\\
 \cline{2-3}
         & \# GePEToS &  \# Data \\\hline 
r = 0 cm & 4,2 mm  &  4,4  mm $^{1}$                 \\\hline 
r = 5 cm & 5,5 mm  &  5,2 mm $^{2}$              \\\hline 
r = 10 cm & 5,6 mm &  6,4 mm $^{2}$                 \\\hline 
r = 15 cm & 6,6 mm &  7,00 mm $^{2}$           \\\hline 
r = 20 cm & 7,9 mm &  8,0 mm $^{1}$ / 8,3 mm $^{2}$ 
 \\ \hline\hline 
\end{tabular}
\end{center} 
\caption{\it Comparison of transaxial resolution between GePEToS and experimental data sets from \cite{bib:ECAT}$^{1}$ and \cite{bib:DATA}$^{2}$ .}
\label{tab:resol}
\end{table}

\subsection{A liquid xenon $\mu$PET camera}
\begin{figure}
\centering
\includegraphics[width=3.0in]{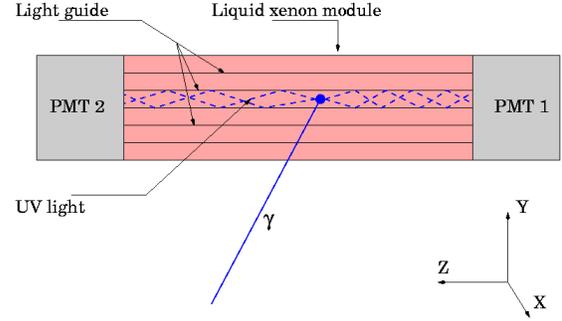}   
\caption{ Sketch of an elementary module of the LXe $\mu$PET camera : 
the z-axis is along the axial direction of the $\mu$PET.}
\label{fig:Module}
\end{figure}

For several years, liquid xenon has been considered by two groups to build PET
cameras~\cite{bib:LXe}~\cite{bib:LXe1}. We have used GePEToS to optimize the design of a
small animal $\mu$PET camera which would exclusively 
use the scintillation light of LXe. 
The active part of the camera is a ring featuring an internal diameter of 
10~cm and a radial extension of approximately 25~mm. It is filled with liquid xenon and placed 
in a cryostat composed of thin aluminum walls (especially around the imaging port).
16 identical modules of the type shown on figure~\ref{fig:Module}, are immersed in this ring. Each module presents a 2 x 2 cm$^2$ cross-section in the transaxial plane of
the camera. The axial field of view spans 5~cm. A module is optically subdivided by 100 2 x 2 mm$^2$ MgF$_2$-coated 
aluminum UV light guides. The UV light is collected on both sides of a module by two 
position sensitive photo-tubes. The (x,y) positions measured by the photo-tubes  
determine which light guides have been fired : hence we measure the
transaxial Depth Of Interaction (DOI) of the photons~\cite{bib:phd}. For each module, the axial coordinate is provided by the following ratio of the photo-tube signals :
$$
\frac{PMT_{1}-PMT_{2}}{PMT_{1}+PMT_{2}}
$$

\begin{figure}
\centering
\includegraphics[width=3.5in]{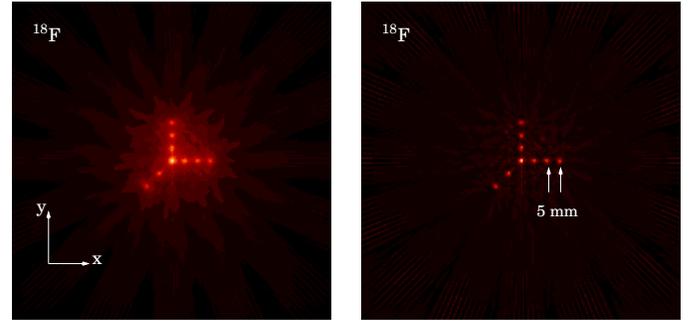}  
\caption{ Reconstructed images of point-like $^{18}$F sources placed in the z=0
transaxial plane of a 4~cm diameter water cylinder . Left : unfiltered ; Right : filtered.} 
\label{fig:Image}
\end{figure}

The performance of this device has been fully evaluated using 
GePEToS plus a dedicated light collection program written in C++. It assumes
a quantum efficiency of the photo-tubes of 15\% and a UV reflection coefficient
of 90\% for the light guides. The simulated $^{18}$F sensitivity of this device evaluated on a water cylinder
of 4~cm in diameter and 4~cm in length is 0.6\% for an energy threshold of 
250~keV. Its image resolution after filtering is 1.6~mm (FWHM) throughout the view field, thanks to the DOI capability of the device. 
Figure~\ref{fig:Image} shows the reconstructed image of point-like $^{18}$F sources placed in the z=0 transaxial plane of 
a 4~cm diameter water cylinder. After filtering, these point-like sources are clearly resolved. Table~\ref{tab:reso} shows the image
resolution at different points in the Field Of View
(FOV) for the $^{18}$F, $^{11}$C and $^{15}$O : we see that the DOI measurements provide a good resolution uniformity in the FOV.

\begin{table}[htbp]
\begin{center}
\begin{tabular}{|c|c|c|c|} \hline
         & $^{18}$F & $^{11}$C & $^{15}$O \\\hline\hline

 r=0 mm & 1,6 mm &2,0 mm  & 3,1 mm \\ \hline
 
 r=5 mm & 1,9 mm & 2,6 mm & 4,0 mm  \\ \hline

 r=10 mm & 1,8 mm & 2,3 mm & 3,7 mm \\ \hline 
 
 r=15 mm & 1,6 mm & 2,3 mm& 3,3 mm\\ \hline \hline 
\end{tabular}
\end{center} 
\caption{\it Transaxial image resolution (FWHM) for point sources at four radius values in the FOV.}
\label{tab:reso}
\end{table}

\subsection{Magnetic field and image resolution}
\begin{figure}
\centering
\includegraphics[width=2.7in]{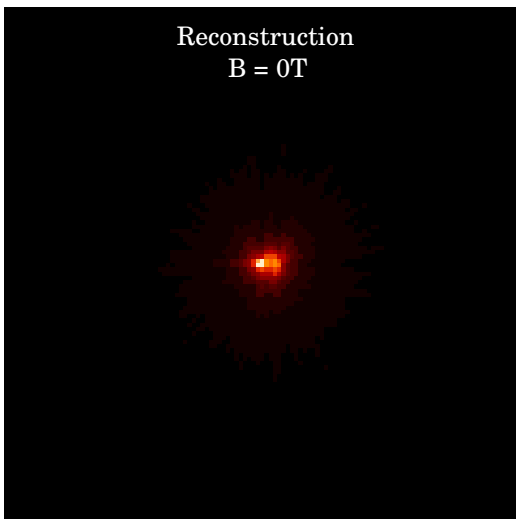}  
\caption{ Reconstructed image of two point-like $^{15}$O sources placed in the z=0
transaxial plane of a 4~cm diameter water cylinder imaged by a Concorde P4 
$\mu$PET. There is no magnetic field and the sources are not separated.} 
\label{fig:ImageBField1}
\end{figure}

\begin{figure}
\centering
\includegraphics[width=2.7in]{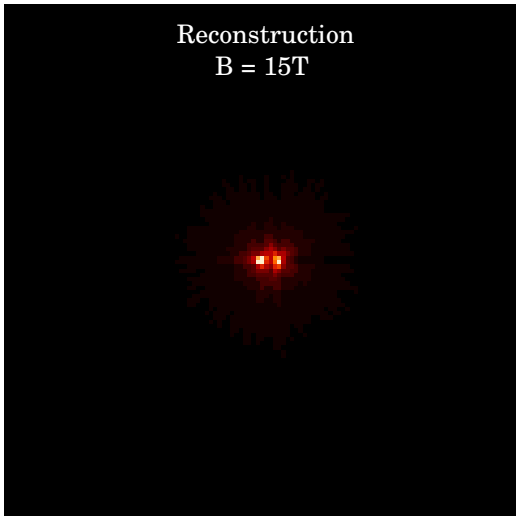}  
\caption{ Reconstructed image of two point-like $^{15}$O sources placed in the z=0
transaxial plane of a 4~cm diameter water cylinder imaged by a Concorde P4 
$\mu$PET. Sources are 4~mm apart.} 
\label{fig:ImageBField2}
\end{figure}

As Geant4 presents the capability to transport and track charged particles 
in strong magnetic fields, we used GePEToS to evaluate the potential image resolution
gain of a P4 Concorde $\mu$PET camera~\cite{bib:P4} which would be operated in the strong 
axial field of a MRI scanner. We found no improvement for $^{18}$F and a marginal
gain for $^{11}$C. However for $^{15}$O and providing the device is operated in 
a 15~T solenoidal field, figures~\ref{fig:ImageBField1} and~\ref{fig:ImageBField2} show that the magnetic field allows to 
resolve two point-like sources separated by 4~mm. The physical explanation  
of this effect is very comparable
to what happens in a TPC (Time Projection Chamber) operated in a magnetic field. The axial magnetic field 
blocks the transaxial diffusion of electrons and confines them within a spiral around their creation 
points. Our results are in good agreement with what was found in previous studies~\cite{bib:BField,bib:BField1}.
15~T MRI scanners are now becoming available for small animals, but the operation of photo-sensors in 
such a strong magnetic field remains a very difficult challenge for the future. 

\section{Conclusion and perspectives} 

We have established the basis of a PET simulation package (GePEToS) based on Geant4 as a transport and tracking engine. GePEToS has 
been validated against the available data of a Siemens  ECAT EXACT HR+ PET scanner. We used GePEToS to guide our development effort toward a LXe PET
camera dedicated to the small animal imaging. The sources of GePEToS can be freely downloaded from this site~\cite{ouaib:gepetos}.  

\section*{Acknowledgments}
This work was made possible thanks to the financial grants allocated by the 
Rh\^{o}ne-Alpes region through its "Emergence" science program and by CNRS/INSERM
via its IPA program dedicated to the small animal imaging. We are also 
indebted to Jean-Fran\c{c}ois Le Bas and Daniel Fagret of the Medical 
department of the Joseph Fourier University of Grenoble for the support and
motivation they brought to this project.


\begin{thebibliography}{1}

\bibitem{bib:P4} A.F. Chatziioannou {\it et al.}, "Performance evaluation of
microPET : a high-resolution oxyorthosilicate PET scanner for animal 
imaging," {\it Journal of Nuclear Medicine}, vol. 40, p. 1164, 1999. 

\bibitem{bib:G4} Geant4 web page : http://wwwinfo.cern.ch/asd/geant4/geant4.html

\bibitem{table}E. Browne, R.B. Firestone, {\it "Table of Radioactive Isotopes."} Virginia S. Shirley Editor. 

\bibitem{posit}P. Colombino {\it et al.}, "Study of positrunium in water and ice from 22 to -144 $^{o}C$ by annihilation
quanta measurements," {\it Nuovo Cimento}, vol. XXXVIII, no 2, 1965.

\bibitem{bib:NIST} NIST web page : http://physics.nist.gov/PhysRefData/

\bibitem{bib:ROOT} ROOT web page : http://root.cern.ch

\bibitem{bib:IDL} IDL web page : http://www.rsinc.com/idl/index.asp

\bibitem{bib:ECAT} G. Brix {\it et al.}, "Performance Evaluation of a whole-body 
PET scanner using the NEMA protocol," {\it Journal of Nuclear Medicine}, vol. 38, p. , 1997. 

\bibitem{bib:DATA} Private communication of Service Hospitalier Frederic Joliot - CEA -
DSV - Orsay France. 

\bibitem{bib:NEMA} National Electrical Manufacturers Association.{\it NEMA
Standards Publication NU 2-1994: Performance Measurements of Positron Emission
Tomographs.} Washington, DC: National Electric Manufacturers Association; 1994. 

\bibitem{bib:LXe} V. Chepel {\it et al.}, "Performance study of liquid xenon detector
for PET," {\it Nucl. Instr. and Meth.}, A392, p.427, 1997.

\bibitem{bib:LXe1}J. Collot, S. Jan and E. Tournefier, "A liquid xenon PET camera for 
neuro-science," {\it IX Int. Conf. On Calorimetry in Part. Phys. - Annecy 2000,} 
Frascati Physics Series, vol. XXI, 305, 2001.  

\bibitem{bib:phd} S. Jan, ph.D. Thesis, University of Grenoble (UJF, Grenoble, France), september 2002.

\bibitem{bib:BField} R. Raylnan {\it et al.}, "Combined MRI-PET Scanner: A Monte Carlo Evaluation of the Improvements in PET Resolution Due to
the Effects of a Static Homogeneous Magnetic Field," {\it IEEE Transactions on Nuclear Science}, vol. 43, p.2406-2412, 1996. 

\bibitem{bib:BField1}  B. Hammer {\it et al.}, "Measurement of Positron Range in Matter in Strong Magnetic Field," {\it IEEE Transactions on Nuclear Science}, vol. 42, p.1371-1376, 1995.

\bibitem{ouaib:gepetos} GePEToS web page : http://isnwww.in2p3.fr/tep/gepetos.html


\end{thebibliography}
\end{document}